# The healing strategy by prioritizing minimum degree against localized attacks on interdependent spatially embedded networks

Kai Gong[1,2,3*], Jia-Jian Wu[1], Qing Li[1], Yi-Xin Zhu[4]

1) (School of Economic Information Engineering, Southwestern University of Finance and Economics, Chengdu 611130, China)
2) (Laboratory for Financial Intelligence and Financial Engineering, Southwestern University of Finance and Economics, Chengdu 611130, China)
3) (Collaborative Innovation Center for the Innovation and Regulation of Internet-based Finance, Southwestern University of Finance and Economics, Chengdu 611130, China)
4) (School of Computer Science and Engineering, Xinjiang University of Finance and Economics, Urumqi 830012, China)

**Abstract**

Many real infrastructure networks, such as power grids and communication networks, are not only depend on one another to function, but also embedded in space. A lot of works have been devoted to reveal the vulnerability of interdependent spatially embedded networks considers random failures. However, recently research show that they are susceptible to geographically localized attacks or failures cased by natural disasters or terrorist attacks, which affect all nodes within a given radius. In particular, small localized attacks may lead to catastrophic consequences. As a remedy of the collapse instability, one research introduced a dynamic healing strategy and the possibility of new connectivity link with a probability, called random healing, to bridge two remaining neighbors of a failed node. The random strategy is straightforward. Here, unlike previous strategy, we propose a simple but effective strategy, called healing strategy by prioritizing minimum degree(HPMD), which establishing a new link between two functioning neighbors with minimum degree during the cascading failures. In addition, we consider the distance between two nodes to avoid change the network structures. Afterwards, a comparison is made between HPMD and three healing strategies: random, degree centrality and local centrality to identify which a pair of neighbors have high priority. Simulation experiments are presented for two square lattices with the same size as interdependent spatially embedded networks under localized attacks. Results demonstrated that HPMD strategy has an outstanding performance against localized attacks, even when ratio of healing is vary. Moreover, HPMD is more timely, more applicability and costless. Our method is meaningful in practice as it can greatly enhance the resilience and robustness of interdependent spatially embedded networks against localized attacks, and eradicate catastrophic collapse in spatial system.

**keywords**: interdependent spatially embedded networks, localized attacks, cascading failures, healing strategy

**corresponding email**: gongkai1210@swufe.edu.cn

## 1. Introduction

It is increasingly clear that almost all real infrastructure networks interact with one another[1]. This has led to an emerging new research field in complex networks that is called interdependent networks[2]. Those interactions keep systems functional, but increase the vulnerability of interdependent networks under random failures or malicious attacks[3]. Because of dependency of nodes in interdependent networks, the failures of a small fraction of nodes will cause the breakdown of the whole system[4]. Thus, more studies on the robustness of interdependent networks have been proposed in Refs[5-8]. However, these studies mainly focus on the interdependent networks in which space restrictions are not considers. In fact, many modern infrastructure networks are embedded in space, the dependency are restricted by their spatial lengths[9]. To model spatial networks, many researchers used 2D lattices[10]. For example, Li et al. [11] introduced a model composed of two square lattices interdependent within a certain dependency length, called *interdependent spatially embedded networks*. Along this pioneering work, several works have been followed for understanding the vulnerability of interdependent spatially embedded networks considers random failures, and found that they are more vulnerable than non-embedded networks[12,13]. It is noteworthy that *localized attacks* on some networks are significantly more damaging than random failures[14-17].

Localized attacks are geographically attacks induced by natural disasters(e.g. earthquakes) or malicious attacks(e.g. burst of atom bomb), i.e., a node is affected, then their neighbors and so on[18, 19]. In short, the main difference between localized attacks and random attacks is that the former always limited to a local area, while the latter are global attacks and the failures are distributed throughout the whole system. Recently research reveals that failures in systems are often not random, but geographically localized attacks[20]. Moreover, such localized attacks results in an extreme instability of interdependent spatially embedded networks since a small failures can trigger an avalanche[21]. Thus, a pressing issue arises as how to enhancing resilience of interdependent spatially embedded networks against localized attacks. In the past years, prevent and mitigation strategies have been proposed to restrain the cascading of failures in interdependent networks[22], including making the number of autonomous nodes with large degree[23,24], protecting the node of high degree as critical node[25,26]. Among them, until recently, the spontaneous recovery mechanism of single networks[27] and interdependent networks, have attracted more attention. Muro et al. [28] showed that recovery mechanism can greatly enhance the resilience of interdependent networks, by recovering the mutually boundary nodes during the cascading of failures and connecting them to the giant components become active again. The recovery strategy in interdependent spatially embedded networks may be suitable to against random failures[29], but not for localized attacks. Because this failure type always simultaneously destroy multiple adjacent devices, including nodes and their edges. In order to avoid catastrophic events, it is necessary to improve the ratio of nodes restored in interdependent spatially embedded networks, and is limited by high cost. On the other hand, when localized attacks occurs considerable effort is made to reorganize the network among existing nodes by *healing*[30], i.e., bridge some remaining nodes by add new links.

A rich body of link addition strategies have been proposed both in single network[31] and interdependent networks[32-35]. For instance, Stippinger et al.[30] develop a dynamic healing model for the competition between the cascading failures and the healing that bridge functioning neighbors of failed node by add new connectivity links. Based on healing model, the authors propose a simple strategy: randomly chosen two remaining neighbors of a failed node. However, there are still some problems to be solved. Firstly, the random strategy test the effects of adding new links under random failures, but in the reality, failures may be the results of localized attacks. Secondly, different failures types bring different attack effects and need different healing strategies[36-39]. To the best of our knowledge, so far there is no research on the healing of interdependent spatially embedded networks suffer from localized attacks. In this paper, we propose a more effective method, *healing strategy by prioritizing minimum degree* (henceforth labeled as HPMD), which establishing a new link between two functioning neighbors with minimum degree of a failed node during the healing process. HPMD is clearly different from previous approaches usually take the node of high degree as critical node, and it consider the distance between a pair of two neighbors to avoid change the network structures. Applying HPMD to simulated networks, we completely show that our strategy is more effective than other healing strategies, including random (RH), degree centrality (HPD), and local centrality (HPL), for the given ratio of healing ω, with higher size of mutual giant connected $S$, higher probability of existence of the mutual giant connected $P\infty$, higher critical attack size $rh_c$ where the system collapses, lower peak of the number of iteration steps(NOI) needed to reach the steady state, fewer addition new links during the healing process <$E_h$> and go on. This opens new avenues of research in the study of resilience in embedded spatial systems.

The paper is organized as follows: Section 2 describes the model of interdependent spatially embedded networks, and the concept of localized attacks. Section 3 introduces the healing model. Section 4 represents the HPMD and Section 5 shows the simulation results and corresponding analysis. Section 6 discusses the HPMD. Section 7 summarizes the work.

**2. Network model and failure type**

In this section, we review the interdependent spatially embedded networks model, and briefly introduce the concept of localized attacks.

*2.1 interdependent spatially embedded networks*

According to Ref.[14], the interdependent spatially embedded networks have two distinct features: (a) each node in network is only connected to the nodes in their spatial vicinity; (b) the dependence links between the networks are not random but have a certain dependency length $r$. The parameter $r$ represents the maximum length a node in one network connects to its dependent node in another network. Here, for the sake simplicity and without loss of generality, we generate the network similarly to Ref.[14], as follows: model dependencies between spatially embedded networks by two square lattices A and B of same number of nodes $N = L * L$ with periodic conditions, where $L$ is the linear size of the lattices. In lattice, each node has two types of links: connectivity links and dependency links. Each node is connected with its neighbors in the same network by connectivity links. Each node in lattice A is depend on a node in lattice B which is chosen at random from all nodes within a radius $r$ via dependency links(and vise versa), That is, a node $a_i$ located at $(x_i, y_i)$ in lattice A is coupled with one and only node $b_j$ located at $(x_j, y_j)$ in lattice B with the condition(see Fig.1):

$$|x_i - x_j| \leq r \text{ and } |y_i - y_j| \leq r \qquad (1)$$

In this work, we used the same set of parameter $r$ as in Ref.[24], namely $r = 15$, and the lattice size was $L=100$ ($N=10000$). The value of <$k$> studied here was 3, based on empirical studies of power grids which have found a mean degree of $2.5 \leq$ <$k$> $\leq 3$[40]. The simulation results in this paper were generated in this manner.

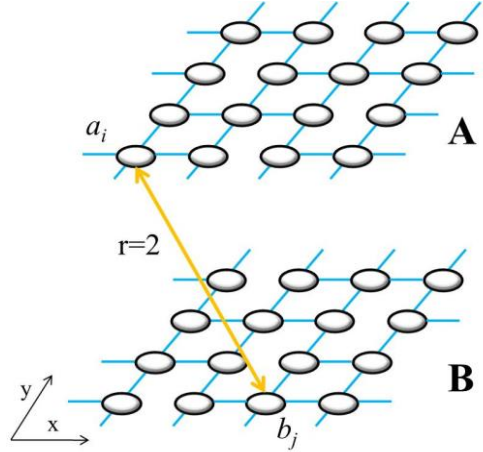

**Fig.1** The interdependent spatially embedded networks was constructed by two square lattices A and B with the same size where each node has two types of links: connectivity links(blue) and dependency links(yellow). Every node initially has up to four neighbors at most via connectivity links, and also, each node $a_i$ located at $(x_i, y_i)$ in lattice A is coupled with one and only one node $b_j$ located at $(x_j, y_j)$ in lattice B via dependency links, with the constraint $|x_i - x_j| \leq r$ and $|y_i - y_j| \leq r$. The dependency link means that two nodes connected by it depend on each other, one of which failed, the other fail too.

*2.2 localized attacks*

Localized attacks model natural hazards or weapons of mass destruction which occur in specific areas. It is a group of failed edges concentrated in a geographical domain, resulting in adjacent isolated nodes. According to the Refs.[24], a local damage forming an initial hole with radius *rh* centered on a randomly node, propagate from a random location to the entire network. Here, following the way introduced in this reference: First, we initiate the localized attacks process by randomly chosen a node in lattice A as a root. Next, we remove all nodes and their edges within a distance *rh* from the root in the lattice. Thus, this triggers a cascade in which the nodes that depend on the removed nodes fail, triggering further losses as more nodes apart from giant component, and triggers further damage due to the dependencies between networks. This process is repeated until no more nodes fail. In the model, as in percolation theory[40], only nodes in the giant component (GC) of the lattice are still functioning.

**3. Healing model**

The healing model assumes a process of healing that is immediately applied at the first stage of the cascading failures to avoiding or delaying the collapse of the interdependent networks[32]. The reason behind this model is based on the fact that (a) failures propagate rapidly in networks, damaged devices can not be timely replaced by new one and (b) in many real infrastructure networks it is reasonable to reinforce adjacent nodes of failed nodes. The rules of the model is as follows: After the initial attacks occurs in network A, the coupled nodes of the failed nodes are removed from the network B through the dependent links, as the conventional process of cascading failures introduced in Ref.[2], but before spreading the failures back to network A the healing step intervenes at that time. The traditional healing step means all pairs of neighbors of each failed node is *randomly* considers as a candidate for a new connectivity link, with the number of add links given by the ratio of healing ω, after having selected the candidates, all new connectivity links are established simultaneously. Due to the dependence of embedded spatial systems, further failures might propagate back and forth within the system, and always intervened by the healing step. Notice, *random healing* may change the topology considerably, bridging distances over time. In this paper, we denote by $n=0,1,…$ the time steps of the cascading process, and the procedure for any stage *n* are given by (see Fig.2):

- Stage *n* in A:

(1) Nodes in lattice A become inactive if they lose dependent nodes in lattice B at state $n-1$.
(2) The remaining nodes in lattice A fail if they do not belong to the $GC_A$ via connectivity links.

- Healing stage in A:

(3) Randomly select a pair of remaining neighbors of a failed node in $GC_A$, and build a connectivity link between them. Self-loop and parallel links are not allowed. This procedure is repeated until the demanded number of new links given by ratio of healing ω.

- Stage *n* in B:

(4) Nodes in lattice B fail if they lose dependent nodes due to the cascade of failures at state *n*.
(5) The remaining nodes are alive in lattice B if they belong to the $GC_B$ via connectivity links.

- Healing stage in B:

(6) Randomly select a pair of remaining neighbors of a failed node in $GC_B$ to establish a connectivity link. Self-loop and parallel links are not allowed. This procedure is repeated until the demanded number of new links given by ratio ω.

This procedure is repeated until a steady state is reached, then we are left with the mutual giant component. Note that in our model a steady state is reached when the system is still functioning and no more nodes fail, or fully collapsed and that there is no intermediate state.

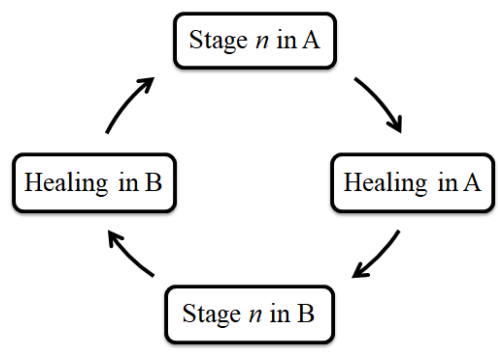

**Fig.2** Schematic healing model on interdependent spatially embedded networks.

## 4. The strategy

In recent study[24], the authors examined the localization of dependency generates a cascading phenomenon which amplifies the local attacks and leads to whole system collapse. When a hole of radius *rh* is removed from the network, the nodes that depended on them must be failed within a distance *r* via dependency links. Then, the secondary damage is highly concentrated around the edge of the hole via connectivity links, leading to the creation of a failure front which propagates outwards, step by step. So, the amount of failure caused per node removed is substantially higher when the failure is localized as compared with random. Based on Section 3, it is quite obvious that the most straightforward healing strategy is random, where two remaining neighbors of fail node are establishing a new link randomly. However, by doing this there is no guarantee that this order of adding links leads to the optimum effect when all the possible sequence of adding links are combined. It may happen that some links which lead to optimal healing have to be established first. Furthermore, random healing might change the lattice structures because it does not consider the distance between two nodes for a new link. In fact, the low degree nodes around the edge of the hole has more damage on the cascading failures, because these nodes are more easily apart from the giant component, failed node can disable corresponding dependent node in other network, and maintain failures propagating[42]. On the other hand, the higher the degree of node, the lower the probability of fail in next time step. Thus, our approach, namely healing strategy by prioritizing minimum degree(HPMD), is reasonable to establish a new connectivity link between remaining neighbors with minimum degree of fail nodes, that is enhancing the connectivity of low degree neighbors and reducing the risk of coupling effect, in order such that the network regains the largest network functionality in each time step. HPMD is quite different from previous approaches usually take the node of high degree as critical node[26-29], and limit the distance between two neighbors of fail node to make sure the connectivity links remain local, conserving the ordered lattice topology. The procedure of this approach is taken after every healing stage in network as follows(see Fig.3):

(1) In healing stage, the set $F_n$ is an union of all failed nodes within sub-network.
(2) Let $L_n$ be the union of all pairs ($v_i$, $v_j$) of remaining neighbors of nodes in $F_n$, self-loop and parallel links are not allowed. Then, we calculate the healing priority for each pairs ($v_i$, $v_j$) during the healing stage. The healing priority index $H(v_i, v_j)$ of node $v_i$ and $v_j$ is defined as

$$H(v_i, v_j) \underset{e_{i,j} \in Vicinity}{=} k_c(v_i) + k_c(v_j) \qquad (2)$$

where $e_{i,j} \in Vicinity$ denotes the link between node $v_i$ and $v_j$ whose shortest paths are no more than 2 in original network, that is all new links were established within a max distance=2. The $k_c(v)$ denotes the number of remaining connectivity links connected to node $v$ in network.

(3) We sort all pairs in the $L_n$ by healing priority index $H$ in ascending order, if some pairs have the same $H$, randomly sort them.
(4) Next, we build a queue $Q = \{H_1, H_2, H_3...\}$. Get node $v_i$ and $v_j$ in a pair from the head of the queue, connect node $v_i$ to node $v_j$ by connectivity link, and repeat this step until the demanded number of new links given by the ratio of healing ω. If the queue is empty, jump to Step 5.
(5) End.

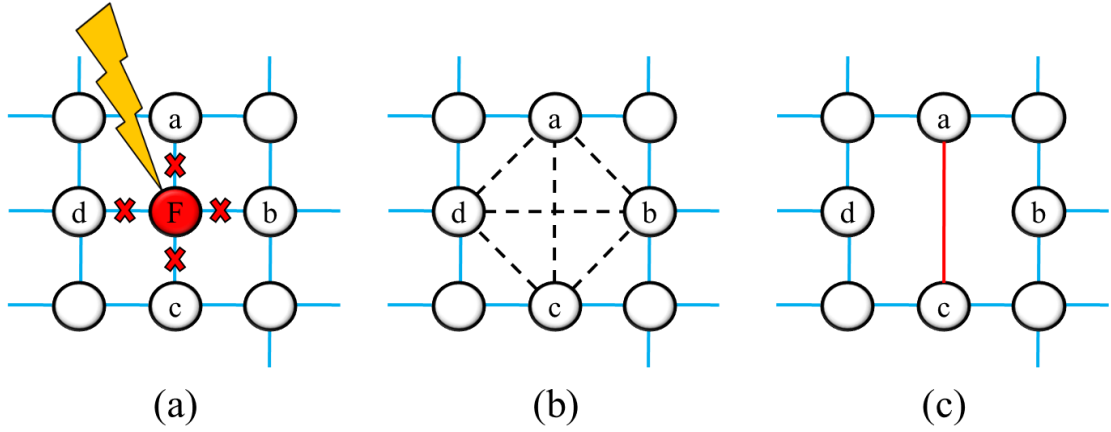

**Fig.3** Schematic illustration of the HPMD: (a)Failure node F that all connections within the GC are removed, and represented by red dot at the end of the arrow, affects its functioning neighbors($v_a,v_b,v_c,v_d$) via connectivity links(blue). (b)In random healing, two remaining neighbors of the fail node F were randomly selected to establish a new connectivity link. (c)According to the HPMD, by counting the number of connectivity links of remaining nodes $v_a$ and $v_c$ ($k_c(v_a)= k_c(v_c)=2$) of the fail node F, and their distance is equal to 2 in original network, after calculating the healing priority index $H(v_a, v_c)=4$, then two remaining neighbors $v_a$ and $v_c$ preferentially establish a new connectivity link(red) to healing the network.

Through computational complexity analysis, HPMD take on the worst-case run times that go as O($N*lg(N)$). Many criteria based on the complex networks theory have been presented for evaluating the importance of nodes[28]. In this study, we mainly adopt three well-known criteria to measure the priority of link addition for a pair of two remaining neighbors in the healing stage: *random*(RH), *degree*(HPD) and *local*(HPL). In random centrality, this is the simplest strategy where a pair of neighbors of fail node is picked randomly for the healing candidate in each healing stage. In degree centrality, the candidate is highly related to its degree, and it is computed as Ref. [42]. In local centrality, the candidate is determined by its nearest and next nearest neighbors, and it is evaluated as Ref. [44].

### 5. Results

To compare the performance of our proposed strategy over healing mechanism, we present experimental results of different strategies including RH, HPD and HPL, on choosing candidates to add new links in interdependent spatially embedded networks(simulated networks for short) within localized attacks. Details about networks construction and localized attacks model are described in Section 2. All experimental results are obtained by average over 10000 independent trials for each ratio of healing ω and attack size *rh*.

For different ratio of healing ω, the robustness of simulated networks under localized attacks with different healing strategies is shown in Fig.4. Here, we use *rh* as initial attack size from network A, and *S* as the size of the mutual giant component at the end of the cascading failures[14,32]. The higher mutual giant component *S* would indicate the better performance of healing, and this figures show the differences for different *rh* and ω. The pink cross-shaped labeled as None in figures show the result without healing design. It is obviously that, the efficiency of healing mechnism are much better than that of without healing after localized attack, and we can ensure that healing mechnism indeed decrease those nodes suffering failure from localized attacks and improve the resilience of simulated network. Then, we compare our strategy with three others to select pairs at healing stage for ω=3%(a), ω=5%(b) and ω=10%(c), it can be seen that their order of $S$ are HPMD > RH > HPD > HPL, whatever ratio ω. For example, $S_{HPMD}$ is on average (0.47) higher than $S_{RH}$ (0.41), $S_{HPD}$ (0.34), and $S_{HPL}$ (0.31), when ω= 5%. It is interesting that RH obtains higher $S$ than HPD. Because higher degree nodes have a relatively lower probability to apart from giant component during the cascading failure, so healing based on degree cannot suppress the process of cascading, which can fail the whole network. On the other hand, lower degree nodes have a higher risk to apart from network, and if one control the lower degree neighbors which coupled with those nodes with average degree of interdependent failure, then cascading failure reduced by localized attack would eliminate a lot, which means it is wise to select remaining neighbors based on this idea to establishing a new link firstly. The results indicate that HPMD is, in general, more effective to against localized attacks, and it will delay the appearance of collapse behavior as the attack size grows.

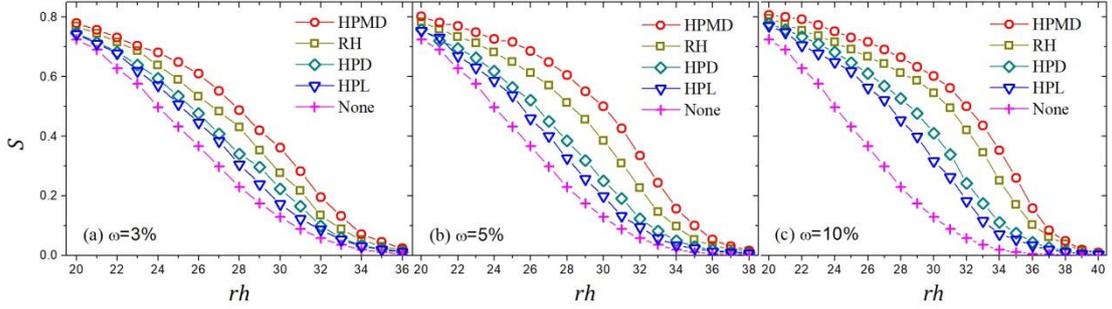

**Fig.4** The size of mutual giant connected $S$ as a function of attack size $rh$ with $r$=15.

Whether a healing strategy is effective in controlling the cascading failures is closely related to its effectiveness in breaking up the network after adding a certain number of new links. Let $P\infty$ [2], with HPMD, RH, HPD and HPLD, be the probability of existence of the mutual giant component by using healing strategies. The higher $P\infty$, the more resilience. In the simulations, for a fixed value of $rh$, we consider the system remain function(that means existence of mutual giant component) after localized attacks if $S$ is more than 0.001 of the realizations and nonfunction(that means collapsed) if $S$ is less than 0.001. Fig.5 shows the results of $P\infty$ for simulated networks with different $rh$, almost all the cases show a similar trend as a function of attack size $rh$ in the size of mutual giant component $S$ in Fig.4. In short, it can be seen that $P\infty_{HPMD}$ is higher than that of $P\infty_{RH}$, $P\infty_{HPD}$ and $P\infty_{HPL}$, and work better than three others in the same condition. For instance, as pictured in the Fig.5(b), even only 5% pairs have reconnected by HPMD, it would get almost average 33% of enhancement on $P\infty$ compared to no healing mechanism, and average 7%(16% and 22%) of enhancement on $P\infty$ to RH(HPD and HPL). In comparison, degree-based(HPD) and local-based healing strategies(HPL) have similar result and both perform worst. From Fig.5(a) and Fig.5(c), we note that HPMD is still a better strategy for different ratio ω has falling slowly show its more advantage. In Fig.4 and Fig.5, both indicate that HPMD is outperforms in a large range of *rh*, that is our strategy can reconstitute on giant component among the surviving nodes, and is more effective to against localized attacks in simulated networks.

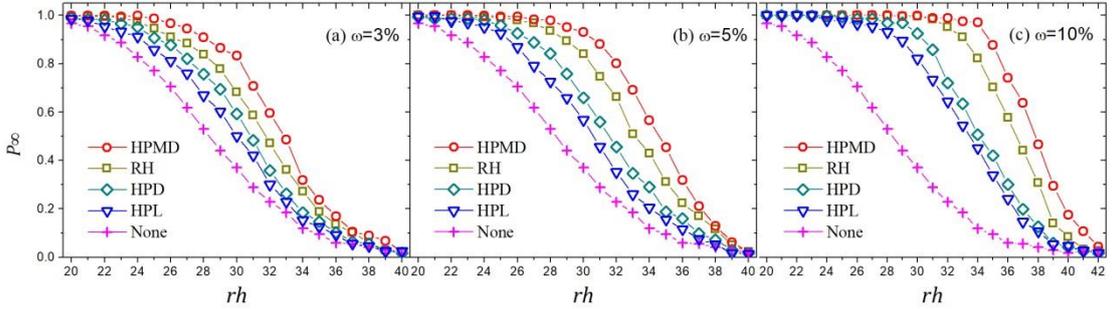

**Fig.5** The probability of existence of the mutual giant component $P\infty$ as a function of attack size $rh$ with $r$=15.

It is an another measure to compare effectiveness of healing strategies, is to observer the number of iteration steps(NOI) in the cascading process, as shown in Fig.6. The NOI is the number of iterative steps needed to reach the steady state which indicates the time scale of the process. It is known that in a conventional cascade of failures without any strategy applied the NOI presents a sharp and peak at the critical threshold[30]. This means that the network requires a long period of time to reach the steady state. Different strategies of healing affects the duration of cascading failures, a good healing strategy is therefore expected to have a low peak of NOI. Fig.6 shows that the peak of NOI is obvious lower for HPMD, as compared with RH, HPD and HPL. For example, in Fig.6(b), the $NOI_{peak}$(HPMD)=17.63 is lower than $NOI_{peak}$(RH)=18.78, $NOI_{peak}$(HPD)=20.33, $NOI_{peak}$(HPL)=21.66 and $NOI_{peak}$(None)=22.63, when ω=5%. From these figure, we can make two observations, (i) that the number of steps enhances as *rh* increases until to reach the steady state, and (ii) that the NOI with HPMD does present a lower peak and steps. The first observation means that as the initial attack size *rh* becomes larger the system requires more NOI to reach the steady state, the second indicates that the required time steps for controlling the failures by HPMD is less at the same condition and show our strategy is more timely.

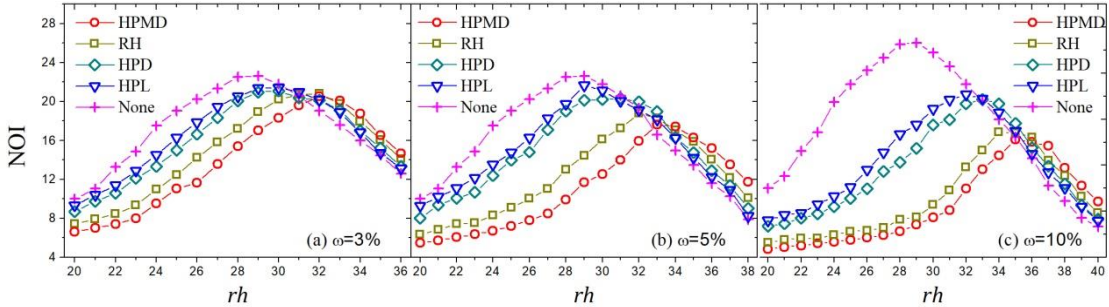

**Fig.6** The number of iteration steps(NOI) in the steady state as a function of attack size *rh* with *r*=15.

According to Ref.[24], depending on two parameters average degree and length of dependency links, the critical attack size $rh_c$ needed for a hole to lead to the collapse of the entire system may be either $rh_c = 0$(unstable), $rhc = \infty$(stable) or $rhc$ =finite (metastable). In stable phase, no matter how large *rh* is (as long as it is finite) the damage will remain localized and the system will stay intact. In unstable phase, the system spontaneously collapses even with *rh*=0 (no localized attack). But, in metastable phase only attacks with radius $rh>rh_c$, it will trigger a cascade which destroys the entire system. If a system suffers a localized attack, what is the difference of $rh_c$ by different healing strategies that break up the networks? The effectiveness of HPMD is further illustrated in Fig.7, which compares the critical attack size $rh_c$ for different healing. Let $rh_c$, larger means more robustness, be the minimum radius of localized attacks needed to cause the system to collapse. Obviously, we can see that $rh_c$ increases when ω increases, the networks become more resilience. The $rh_c$ values used by HPMD is always greater than three others, and is significantly more effective. For instance, when ω=5% in the Fig.7, the $rh_c$(HPMD)=39.54 higher than $rh_c$(RH)=38.88, $rh_c$ (HPD)=37.93 and $rh_c$(HPL)=37.71. Clearly, within a small certain limit ratio of healing, HPMD is the first choice to determine which pairs of remaining neighbors of failed node that would have add a new link firstly.

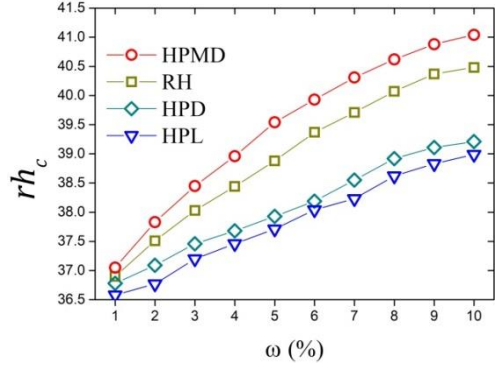

**Fig.7** The critical attack size $rh_c$ as a function of healing probability ω with r=15.

A more efficient healing strategy is one that need fewer new links before network reach the steady state. For each realization of localized attacks with healing, we calculated how many new links were introduced in simulated network until the steady state is reached, resulting to <$E_h$>. Here, let <$E_h$> be the average number of new connectivity links during the healing process until to reaching the steady state. Fig.8 gives <$E_h$> as a function of attack size $rh$ with different ω. From this figure, it is clearly that HPMD adding a relatively fewer new links will obtain a significant enhancement of simulated network while with a tiny cost incrementation. In the same conditions, we can observe that HPD and HPL will have a high cost though they can also improve the resilience of network. For example, <$E_h$>$_{HPMD}$ is on average (146.38) fewer than <$E_h$>$_{RH}$ (164.15), <$E_h$>$_{HPD}$ (186.87) and <$E_h$>$_{HPL}$ (195.95), when ω=5%. From Fig.8(a) and Fig.8(c), HPMD is still have better performance for <$E_h$> values have increasing slowly show its costless. It is a result of the addition new links between remaining functional neighbors with minimum degree of failed node by HPMD, then the healing stage require smaller number of new links to reorganize the remaining nodes. The result further confirms that HPMD can greatly enhance the resilience in a reasonable cost.

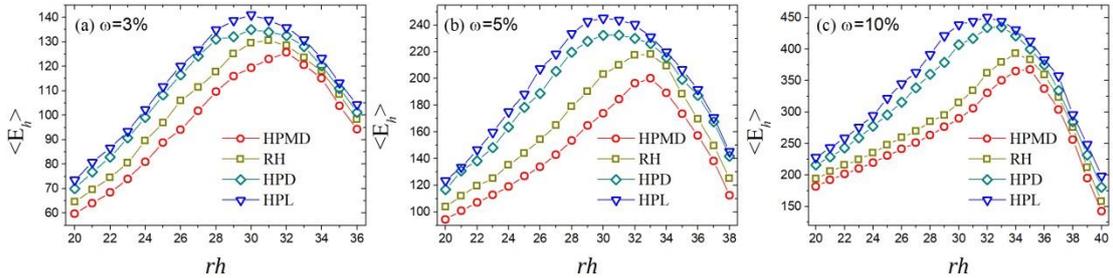

**Fig.8** The average number of established new connectivity links during the healing process <$E_h$> as a function of attack size $rh$ with $r$=15.

It is confirmed that the weakness of healing strategy without consider the distance between nodes, of course, may change the topology considerably, bridging longer distance as the time goes on[32]. Here, the long range connectivity links refers to the link between two nodes whose shortest paths are more than 2 in original network. In reality, two nodes which were far away try to establishing a new link in spatial network, neither rarely practical, nor efficient. In order to further analyze the structural integrity of network by HPMD, Table.1 shows the proportion of long range connectivity links by healing strategies until reach to steady state with ω =5%. Let $d$ be the shortest paths between two nodes in original network. Then, we calculate the distance proportion $P(d)$ of the connectivity links between two nodes in network with $rh$ constant. For example, $P(d=1)$ refers to the proportion of the links between two nodes in remain networks, those shortest path is equal to one in original network. The average of $\bar{P}(d)$ is defined as

$$\bar{P}(d) = \overline{\sum\nolimits_{rh} P(d)} \quad (3)$$

where $rh \in [20,36]$ and delta is 1. In Table.1, when $\bar{P}(d=1)$, it can be obviously seen that their order of $\bar{P}(d=1)$ are HPMD > RH > HPD > HPL. In contrast, when $\bar{P}(d=2)$, the order has been reversed, HPMD < RH < HPD < HPL. It is note worthy that the average distance proportion of the links between distant nodes, $\bar{P}(d=3)$ or $\bar{P}(d=4)$, is equal to zero by HPMD, compared with three others. The reason for this table is simply, because HPMD aims at adding the link only between adjacent remaining neighbors of failed node at each healing step, in the sense that the changing of network structure is minimum. So, the giant component at the end of the cascading process is still lattice structure by HPMD. In general, our proposed strategy can maximum to avoid change the network topology through establishing a new link between adjacent neighbors of failed node, and be more applicability.

|      | $\bar{P}(d=1)$ | $\bar{P}(d=2)$ | $\bar{P}(d=3)$ | $\bar{P}(d=4)$ |
|------|---------------|---------------|---------------|---------------|
| HPMD | 99.46%        | 0.54%         | 0             | 0             |
| RH   | 99.29%        | 0.69%         | 0.01%         | 0.01%         |
| HPD  | 99.05%        | 0.88%         | 0.05%         | 0.02%         |
| HPL  | 98.47%        | 1.39%         | 0.12%         | 0.02%         |

**Table 1.** Comparisons of the average distance proportion of the connectivity links between nodes in the steady state of the mutual giant component by ω=5% and $r$=15, when $rh \in [20,36]$.

## 6. Discussion

Indeed, from all figures and table in Section 5, both illustrates that HPMD is the most effective and costless healing strategy to improving the resilient of interdependent spatially embedded networks against localized attacks. The reasons why our proposed method can reinforce the resilience of the networks are as follows. Firstly, lower degree nodes are more easily apart from giant component, then coupled node in other network would be failed too. Fig.9 is shown as the probability of failure with different degree vary in time steps under localized attacks, without any healing. Except for first step, the probability of failure of degree $k$=1 is on average (48.21%) higher than $k$=2(33.89%), $k$=3(15.05%) and $k$=4(3%), when $rh$=32. It is obvious that, node who less connectivity are more likely to apart from current giant component for each step. Secondly, the coupling in networks will leads to one interdependency link between a low degree node and another node with average degree, thus magnifying the impact of low degree nodes. This is completely difference in a single network, the failures of lower degree nodes not have a great impact on network because the weakness of connectivity. HPMD aims at adding the link between two remaining neighbors with lower degree of failed node at each healing step, in the sense that the increase of network functionality is maximum at each step.

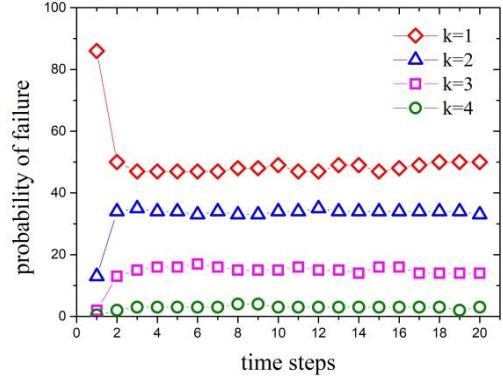

**Fig.9** The probability of failure with different degree in network A, as a function of the finite time steps during the cascading process with *rh*=32.

There is no doubt that the performance of healing strategy depend on network structures. To show the influences of the spatial constraints on healing, we have performed experiments about the critical attack size $rh_c$ as a function of length of dependency links *r* in simulated networks, as shown in Fig.10. Even the length of dependency links *r* is variable, we can see that the $rh_c$ value of HPMD is higher than three others although they may generate not worse healing results. Clearly, HPMD performs well when the spatial lengths of simulated networks are modified. Furthermore, from the Fig.10, we can make three observations: (i) for a small spatial lengths(e.g. $r \leq 10$), the nodes can only couple their adjacent nodes, in these cases, a localized attacks outbreak is mostly restricted in a local area and a larger attack size is required to initiate a cascade. For instance, the $rh_c$ values of healing strategies both peaked at *r*=10, $rh_c$(HPMD)=43.39, $rh_c$(RH)=43.25, $rh_c$(HPD)=43.06 and $rh_c$(HPL)=42.94; (ii) for an intermediate spatial length, the failure could not only propagate far away, but also lead to a large enough density of failed nodes, thus the network becomes most vulnerable such as $rh_c$ reach to the lowest when *r*=30, $rh_c$(HPMD)=34.79, $rh_c$(RH)=34.01, $rh_c$(HPD)=33.32 and $rh_c$(HPL)=33.05. But even so, HPMD is still maintains dominance over the spatial length; (iii) for a large spatial lengths(e.g. *r*=L)that a given node's dependency link can be located farther away, failure can also propagate far away, but the density of failed nodes is too sparse to trigger a cascading process. For example, the $rh_c$ values both have a relatively higher at *r*=100, $rh_c$(HPMD)=38.52, $rh_c$(RH)=38.07, $rh_c$(HPD)=37.46 and $rh_c$(HPL)=37.11. In general, if $r \to \infty$ or $r \to 0$, the avalanche would not happen because the secondary damage would spread everywhere uniformly or remain in local place, respectively. To be sure, HPMD is a better healing strategy and more suitable to against localized attacks on interdependent spatially embedded networks.

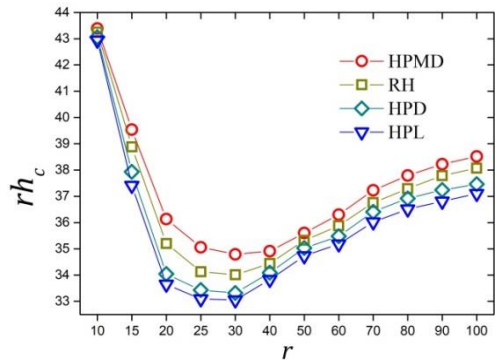

**Fig.10** The critical attack size $rh_c$ as a function of the length of dependency links *r* with ω=5%.

## 7. Conclusions

Robustness of interdependent spatially embedded networks under cascading failure has become hot topic in recent years. Compared with random failures, more serious problem was the damage reduced by localized attacks could make a whole network collapse only need a tiny attack size. A dynamic healing strategy was proposed to determining the consequences of healing by link formation in interdependent networks under random failure. By establishing new random links in remaining neighbors of the failed nodes, the healing strategy delayed the collapse of network through the enhancing of connectivity. However, random healing is too simple, and do not take into account the propagation process of localized attacks in interdependent spatially embedded networks. In this paper, we have analyzed the healing process and localized attacks model, namely the low degree nodes are more easily failed, thus corresponding coupled nodes with average degree would be also disconnected, and maintain failures propagating. Therefore, the remaining functional neighbors with lower degree of failed nodes urgently need healing, and should consider distance between nodes in a new link to avoid changing the network structure. Then, by analyzing these factors, a new healing method HPMD(healing strategy by prioritizing minimum degree) is proposed. In Section 5, to verify the effect of our proposed healing strategy on the optimization of the resilience of the network, we conduct a series of comparative experiments compared with three others (RH, HPD and HPL are conducted in Section 4). It illustrates that HPMD remarkably outperforms others with more efficiency and costless, timely and more applicability. Our experiments also demonstrate that interdependent spatially embedded networks with healing mechanism are more robust to localized attacks than those without healing design.

Our strategy can aid in the development of intervention strategies for crisis situations. Meanwhile, the research of this paper is helpful to provide guidance on how to build robust interdependent spatially embedded networks against the potential localized attacks.

## Acknowledgements

The research is supported by the National Natural Science Foundation of China under Grant 61751110, the Fundamental Research Funds for the Central Universities of Ministry of Education of China under Grant JBK1801080 and JBK170133, the Sichuan Science and Technology Program under Grant 2018JY0607, and the Scientific Research Foundation of the Education Department of Sichuan Province of China under Grant 17ZB0434.